\documentclass[aps,prl,twocolumn,showpacs,showkeys,footinbib,groupedaddress]{revtex4-2}
\usepackage{amsmath}
\usepackage{graphicx}
\usepackage{bm}
\usepackage{color}
\usepackage{epstopdf}

\usepackage{hyperref}
\usepackage[all]{hypcap}

\newcommand{\e}{\ensuremath{\mathrm{e}}}

\begin{document}

\title{Signatures of Topological Transitions in the Spin Susceptibility of Josephson Junctions}

\author{
Joseph D. Pakizer and Alex Matos-Abiague
}

\affiliation{
\textit{Department of Physics \& Astronomy, Wayne State University, Detroit, MI 48201, USA}
}

\date{\today}
\begin{abstract}
We theoretically investigate how the spin susceptibility of a planar Josephson junction is affected when the system transits into the topological superconducting state. We show that the magnetic flux and magnetic field dependence of the spin susceptibility closely maps the phase diagram of the system. In the absence of an external magnetic flux the system can self-tune into the topological superconducting state by minimizing its free energy. Self-tuned topological transitions are accompanied by sharp peaks in the spin susceptibility, which can therefore be used as measurable fingerprints for detecting the topological superconducting state. Away from the peaks, the amplitude of the spin susceptibility can provide qualitative information about the relative size of the topological gap. The signatures in the spin susceptibility are robust, even in junctions with narrow superconducting leads, where critical current minima may no longer serve as an indication of topological phase transitions. 
The predicted results could be particularly relevant for future experiments on realization and detection of the topological superconducting state in planar Josephson junctions.
\end{abstract}

\maketitle


\emph{Introduction}---Topological superconductivity (TS) is a phase of matter supporting the formation of zero-energy bound states, referred to as Majorana bound states (MBSs), which are protected by an energy gap against smooth local perturbations~\cite{Kitaev2001:PU,Alicea2012:RPP,Leijnse2012:SST,Beenakker2013:ARCMP,Aguado2017:RNC}. Since MBS pairs are energy degenerate, exchanges of MBSs from different pairs are equivalent to the application of unitary transformations whose form depends on the topology of the exchange path but not on the path local details~\cite{Nayak2008:RMP,Kitaev2003:AP,Alicea2011:NP}. This allows for the realization of robust qubits and quantum gates with promising applications for fault-tolerant quantum computing~\cite{Nayak2008:RMP,Kitaev2003:AP,Alicea2011:NP}.

MBSs were predicted to naturally emerge in p-wave superconductors \cite{Ivanov2001:PRL,Kitaev2001:PU}. However, TS can also be engineered by using a conventional s-wave superconductor in proximity to a material with nontrivial spin structure, typically provided by spin-orbit coupling (SOC) \cite{Fu2008:PRL,Lutchyn2010:PRL,Oreg2010:PRL,Rokhinson2012:NP,Pientka2012:PRL,Dominguez2017:NPJQM,Fleckenstein2018:PRB,Schuray2020:EPJST} and/or magnetic textures \cite{Klinovaja2012:PRL,Kjaergaard2012:PRB,Fatin2016:PRL,MatosAbiague2017:SSC,Marra2017:PRB,Zhou2019:PRB,Desjardins2019:NM,Mohanta2019:PRA,Steffensen2020:Arxiv}. The
proximity induced superconductivity provides the necessary particle-hole symmetry,
while the SOC makes possible the realization of odd-in-momentum triplet pairing. An
external magnetic field is then applied in order to break the time-reversal symmetry.

The experimental detection of a zero-bias conductance peak (ZBCP) in proximitized nanowires~\cite{Mourik2012:S,Das2012:NP,Deng2012:NL,Deng2016:S,Manna2020:PNAS}, atomic chains \cite{NadjPerge2014:S,Pawlak2016:NPJQI}, and planar Josephson junctions (JJs) \cite{Fornieri2019:N,Ren2019:N,Dartiailh2021:PRL,Hart2014:NP} has provided support for the existence of TS. However, since ZBCPs may emerge even in the absence of TS and the measured ZBCP amplitudes remain well below the predicted universal conductance value of $2e^2/h$ \cite{Sengupta2001:PRB,Law2009:PRL,Flensberg2010:PRB}, the actual origin of the ZBCPs is not yet conclusive. As an alternative to ZBCP measurements, minima in the critical current of JJs, accompanied by phase jumps have been considered
as signatures of transition to the TS state \cite{Pientka2017:PRX,Cayao2017:PRB}. The correct interpretation of the experimental data imposes some challenges because other mechanisms (e.g., Fraunhofer interference, anomalous Josephson effect) may also lead to critical current minima \cite{Kontos2002:PRL,Yokoyama2014:PRB}. Furthermore, it has also been shown that the presence of critical current minima may not necessarily be associated with transitions between the trivial and TS states \cite{Setiawan2019:PRB} and a further analysis of magnetic and crystalline anisotropies of the TS state \cite{Scharf2019:PRB,Pakizer2021:PRR} may be needed to better understand the physical origin of the critical current minima \cite{Dartiailh2021:PRL}. In spite of the challenges, planar JJs continue to attract attention due to their experimental feasibility, versatility, and enhanced region of system parameters supporting the TS state \cite{Hell2017:PRL,Pientka2017:PRX,Setiawan2019:PRB2,Zhang2020:PRB,Woods2020:PRB,Laeven2020:PRL,Zhou2019:PRL,Zhou2021:arxiv,Scharf2019:PRB,Pakizer2021:PRR,Lesser2021:arxiv,Beenakker2019:SPP,Stenger2019:PRB,Svetogorov2021:PRB,Paudel2021:arxiv,Pekker2013:PRL}. Critical current signatures of TS in one-dimensional geometries, like nanowire junctions have also been proposed\cite{San-Jose2014:PRL,Cayao2020:arXiv}.

In this Letter we show that, compared to critical current measurements, the high sensitivity of the spin susceptibility to gap closings provides robust and more direct signatures of the transition of planar JJs to the TS sate. The spin susceptibility exhibits sharp peaks at transitions between trivial and TS states. On the other hand, the spin susceptibility value away from the peaks can be used to identify the magnetic field strength leading to the largest topological gap.

We consider a magnetic field applied in the plane of the JJ. Therefore, magnetic orbital effects can be neglected and the spin susceptibility of the planar JJ becomes proportional to the magnetic susceptibility. The proportionality no longer holds when diamagnetic contributions from the top superconductors become relevant. However, the distinctive peaked behavior of the spin susceptibility at the topological phase transitions will still be reflected in the magnetic susceptibility. The magnetic susceptibility of JJs has previously been investigated both theoretically and experimentally \cite{Auletta1995:PRB,Rae1996:SST,Sergeenkov2008:PLA, Borcsok2019:SR,Araujo-Moreira1997:PRL,Barbara1999:PRB,Araujo-Moreira2004:SSC,Rivera2010:PC}, although not in relation to TS.

\begin{figure}[t]
\centering
\includegraphics*[width=8.5cm]{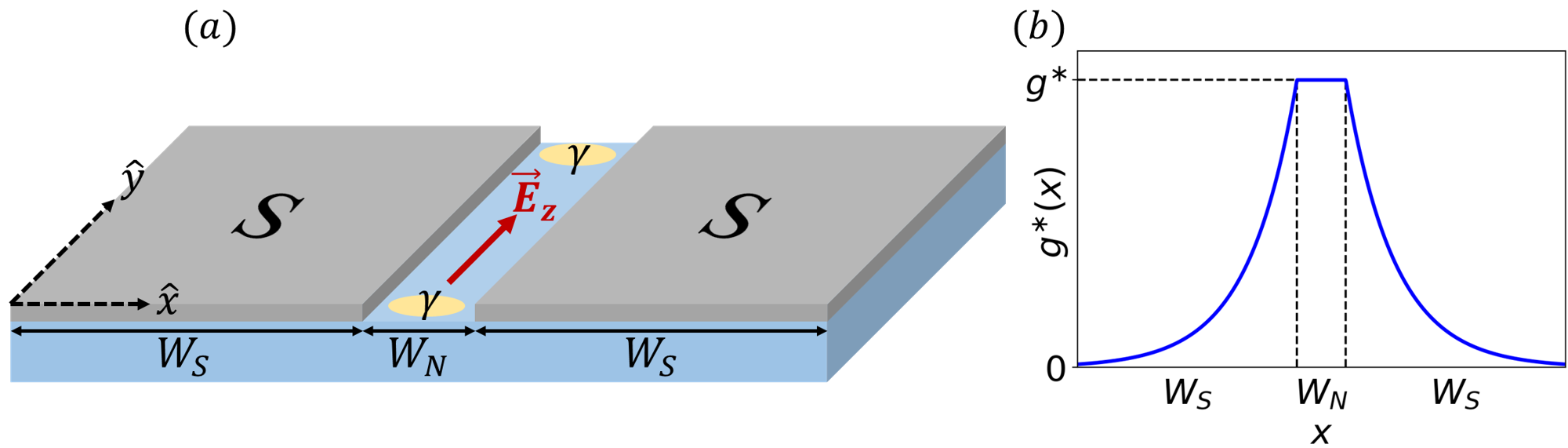}
\caption{(a) A planar JJ composed of a semiconductor 2DEG in contact with two superconducting (S) leads. An external magnetic field is applied along the junction direction. Yellow regions indicate the localization of MBS (labeled by $\gamma$ at the ends of the junction). (b) Position dependence of the effective g-factor, $g^{\ast}$.  The effective g-factor is maximum in the N region and decays in the S regions.
}\label{fig1:syst}
\end{figure}

\emph{Theoretical Model}---We consider a planar Josephson junction (JJ) composed of a 2D electron gas (2DEG) formed in a semiconductor and subject to an in-plane magnetic field $\mathbf{B}$ [see Fig.~\ref{fig1:syst}(a)]. Superconducting regions (S) are induced in the 2DEG by proximity to a superconducting cover, such as Al or Nb, while the uncovered region remains in the normal (N) state. The system is described by the Bogoliubov-de Gennes (BdG) Hamiltonian
\begin{equation}\label{H-BdG}
H=H_{0}\tau_z-\mathbf{E}_Z (x)\cdot\boldsymbol{\Sigma}+\Delta(x)\tau_+ +\Delta^\ast(x)\tau_-\;,
\end{equation}
where
\begin{eqnarray}\label{Ho}
H_{0}&=&\frac{\mathbf{p}^2}{2m^\ast}+V(x)-(\mu_S-\varepsilon)+\frac{\alpha}{\hbar}\left(p_y\Sigma_x - p_x\Sigma_y\right)\;,
\end{eqnarray}
and $\Sigma_{x,y,z}$ and $\tau_{x,y,z}$ represent Dirac spin and Nambu matrices respectively with $\tau_\pm =(\tau_x\pm i\tau_y)/2$. Here $\mathbf{p}$ is the momentum, $m^\ast$ the electron effective mass, $\alpha$ the Rashba spin orbit coupling (SOC) strength, and $V(x)=(\mu_S - \mu_{N})\Theta(W_N/2-|x|)$ describes the difference between the chemical potentials in the N ($\mu_N$) and S ($\mu_S$) regions. The chemical potentials are measured with respect to the minimum of the single-particle energies, $\varepsilon=m^\ast\alpha^2/(2\hbar^2)$.

The second contribution in Eq.~(\ref{H-BdG}), represents the Zeeman interaction. For a magnetic field $B$ applied along the $y$ direction, $\mathbf{E}_Z (x)=[g^\ast(x) \mu_B/2]B(0,1,0)^T$,
Here $\mu_B$ is the Bohr magneton and $g^\ast(x)$ is the position-dependent effective g-factor, which may decay in the S-proximitized regions. Here we consider an effective g-factor with a position dependence given by,
\begin{equation}\label{def-f(x)}
g^\ast(x)= \left\{
\begin{array}{ll}
      g^\ast e^{-|W_S-x|/\gamma} & x\leq W_S \\
      g^\ast & W_S < x\leq W_S+W_N \\
      g^\ast e^{-|W_S+W_N-x|/\gamma} & x > W_S+W_N \\
\end{array} 
\right.,
\end{equation}
where $g^\ast$ is the effective g-factor in the N region and $\gamma$ represents the Zeeman field penetration length characterizing the decay of the effective g-factor in the S regions [see Fig.~\ref{fig1:syst}(b)]. Previous works have considered the case where $\gamma \rightarrow 0$, in which the effective g-factor is finite only in the N region \cite{Pientka2017:PRX,Ren2019:N,Fornieri2019:N,Scharf2019:PRB}, or the limit $\gamma \rightarrow \infty$, in which the effective g-factor is finite and constant over the whole structure \cite{Pientka2017:PRX,Setiawan2019:PRB,Dartiailh2021:PRL,Pakizer2021:PRR}). In what follows we use $E_Z=(g^\ast \mu_B/2)B$ to denote the Zeeman energy in the N region. The spatial dependence of the superconducting gap is $\Delta(x)=\Delta \e^{i\,{\rm sgn}(x)\phi/2}\Theta(|x|-W_N/2)$, where $\phi$ is the phase difference across the JJ.

The spin susceptibility tensor is given by $\chi_{ij}=\partial\langle S_i\rangle/\partial B_j$, where,
\begin{equation}\label{Spin}
\langle S_i\rangle=\frac{1}{Z}\sum_{n}\frac{\hbar}{2}\langle n|\Sigma_i e^{-\beta H}|n\rangle.
\end{equation}
Here $Z=\sum_n \langle n|e^{-\beta H}|n\rangle$ denotes the partition function, $\beta=1/(k_B T)$ and $|n\rangle$ are the eigenstates of the BdG Hamiltonian.

\begin{figure}[t]
\centering
\includegraphics*[width=8.5cm]{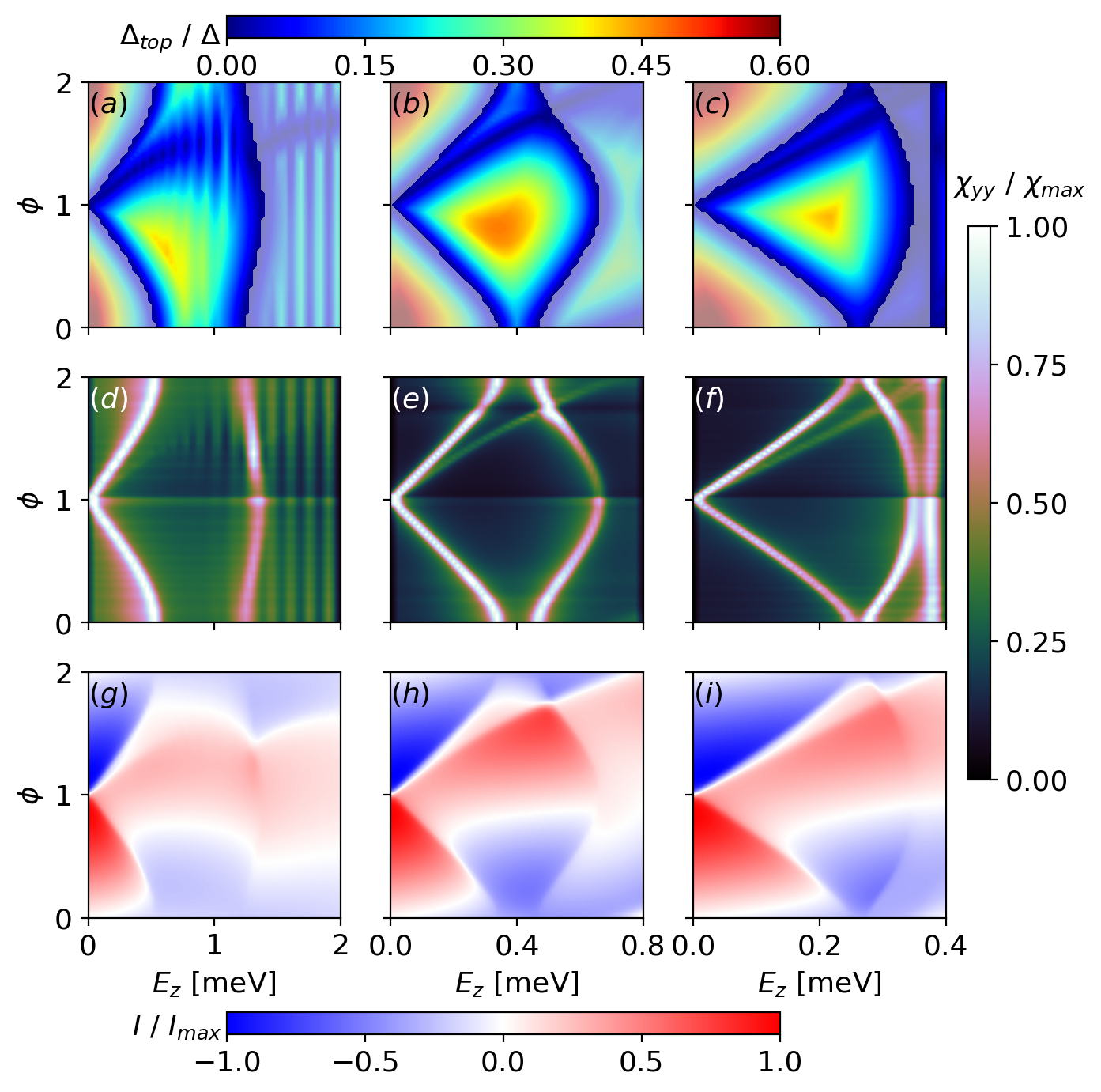}
\caption{Top row: Topological gap as a function of the Zeeman energy, $E_Z$ and the phase difference $\phi$ for penetration lengths $\gamma=1$~nm (a), $\gamma=10^2$~nm (b), and $\gamma=10^4$~nm (c). The gray-shaded and non-shaded areas correspond to topological charge $Q=1$ (trivial state) and $Q=-1$ (TS state), respectively. Middle row: Same as in top row but for the spin susceptibility, $\chi_{yy}$ normalized to $\chi_{max}$ (the maximum of the spin susceptibility at each value of $\gamma$ and $E_z$). Bottom row: Same as in top row but for the supercurrent, $I$, normalized to its maximum value $I_{max}$. The following parameters were used: $W_S=450$~nm, $W_N=100$~nm, $a=20$~nm, $\mu_S=1$~meV, and $\mu_N=0.7$~meV. The other system parameters are specified in the text.
}\label{fig2:ic-top}
\end{figure}

Since the magnetic field is applied in the $y$ direction, we focus on the $yy$-component of the spin susceptibility tensor. After some mathematical manipulations (see the Supplemental Material for details \cite{SM}), we obtain $ \langle S_y\rangle = -(\hbar/2)\partial F/\partial E_Z$,
and
\begin{equation}\label{chi-ez}
    \chi_{yy}=-\left(\frac{\hbar}{2}\right)\left(\frac{g^\ast \mu_B}{2}\right)\frac{\partial^2F}{\partial E_Z^2}
\end{equation}
where $F=-k_B T\ln{Z}$ is the free energy, which in the zero temperature limit reduces to
\begin{equation}\label{free-0}
    F=\sum_{E_n<0}E_n=-\frac{1}{2}\sum_{n}|E_n|.
\end{equation}

We use a tight-binding Hamiltonian resulting from the finite-difference discretization of Eq.~(\ref{H-BdG}) \cite{SM} to compute the energy spectrum, topological gap, ground state phase, critical current, and spin susceptibility. The numerical simulations of the tight-binding version of the BdG Hamiltonian were performed by using the Kwant package \cite{Groth2014:NJP}. We consider Al/HgTe JJs with the following parameters: $m^\ast=0.038~m_0$ (with $m_0$ the bare electron mass), $\Delta = 0.25$~meV, and $\alpha=16$~meV~nm. We consider the three values, $\gamma=1,10^2,10^4$~nm, which correspond to no, partial, and total penetration, respectively, of the Zeeman interaction into the S region.

\emph{Phase-Biased JJs}---In a phase-biased JJ, the phase difference between the two S regions is fixed by an external magnetic flux. The topological gap (i.e., the finite eigenenergy closest to zero) protecting the MBS, the spin susceptibility, and the supercurrent are shown in the top, middle, and bottom rows of Fig.~\ref{fig2:ic-top}. The left, central, and right columns correspond to $\gamma=1,10^2,10^4$~nm, respectively. The gray-shaded and non-shaded areas in (a)-(c) represent trivial (i.e., with topological charge $Q=1$ ) and TS phases (i.e., with topological charge $Q=-1$) \cite{SM}, respectively. As shown in Figs.~(\ref{fig2:ic-top})(a)-(c), when the penetration of the Zeeman interaction into the S regions increases, the phase diagram dependence on the Zeeman field becomes stronger, requiring smaller fields to realize the TS state.

A remarkable observation from Fig.~\ref{fig2:ic-top} is that, unlike the supercurrent (bottom row), the spin susceptibility (middle row) greatly resembles the inverse monotonic behavior of the topological phase diagram (top row).
The white-pink contours in Figs.Fig.~\ref{fig2:ic-top}(d)-(f) corresponding to spin susceptibility peaks ( SSPs) indicate topological phase transitions at gap closings where the fermion parity of the system changes. Furthermore, the spin susceptibility can also capture the qualitative behavior of the topological gap, with smaller amplitudes indicating larger topological gaps. This makes the spin susceptibility a very promising measurable quantity for experimental detection of topological phase transitions in planar JJs.

\begin{figure}[t]
\centering
\includegraphics*[width=8.5cm]{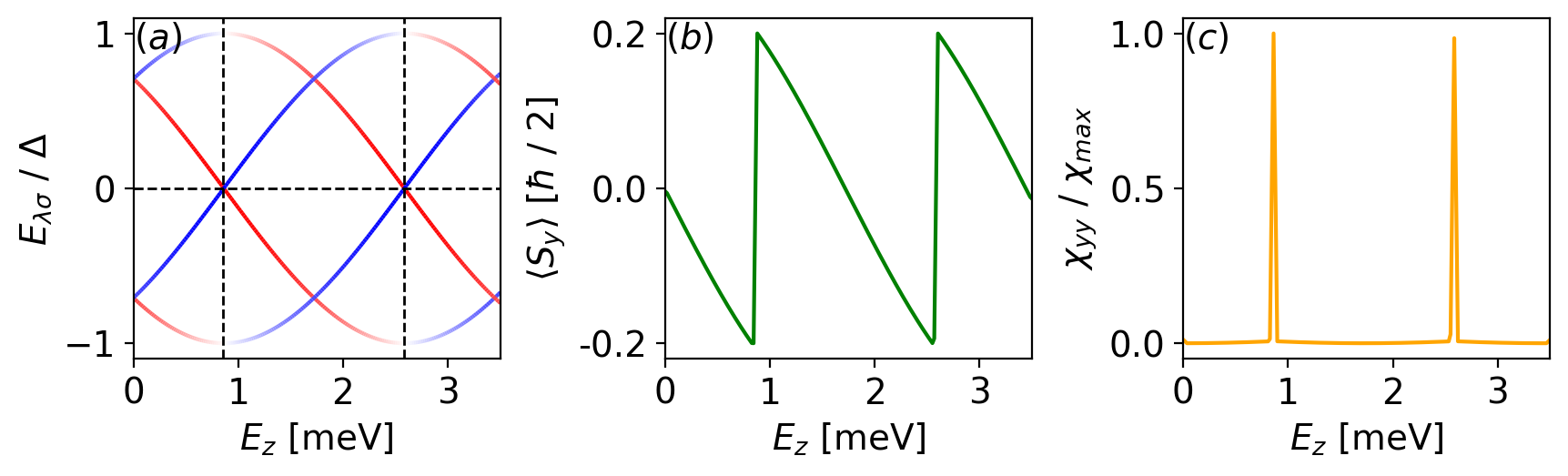}
\caption{(a) Energy spectrum at $k_y=0$ and $\phi=\pi/2$ as a function of the Zeeman energy. The line color indicates the expectation value of the $y$-component of the spin, from antiparallel 
(blue) to parallel (red) to the magnetic field. The vertical dashed lines indicate the topological transitions.
(b) Average spin projection along the $y$ axis for $k_y=0$, and $\phi=\pi/2$. 
The spin susceptibility in (c) has been normalized to its maximum value, $\chi_{max}$. The results were obtained by using a simplified analytical model \cite{SM} with the following parameters: $W_N=100$~nm, $a=5$~nm, and $\mu_S=\mu_N=3$~meV.
}\label{fig3:biased}
\end{figure}

To better understand the high sensitivity of the spin susceptibility to gap closings, we use a simplified analytical model for the subgap states of a JJ with translational invariance along the $y$-axis \cite{SM}. The energy spectrum as a function of $E_z$ is shown in Fig.~\ref{fig3:biased}(a) for the case of a JJ with Zeeman interaction only present in the N region and a fixed phase $\phi=\pi/2$. The vertical dashed lines indicate the Zeeman energies at which the fermion parity of the system changes. The first and second parity changes signal transitions from trivial to topological and back to trivial phases. Line colors denote the spin character of a given state, i.e., the expectation value of the $y$-component of the spin, from antiparallel (blue) to parallel (red) to the field. When the gap closes, the spin character of the energy band closest to zero energy changes abruptly \cite{SR}. This results in jumps in the total average spin, as shown in Fig.~\ref{fig3:biased}(b). The jumps in the average spin produce SSPs [see Fig.~\ref{fig3:biased}(c)] at values of $E_Z$ at which topological phase transitions occur. 

This behavior is \emph{universal}, in the sense that it is independent of the specific model, level of approximation, and junction. Indeed, the inclusion of other effects (e.g., self-consistent treatment of the interrelation between magnetic field and superconducting gap, magnetic field dependent phase gradient, etc) may lead to changes in the topological diagram, but the spin susceptibility will still exhibit a peak any time the spin average jumps due to the occurrence of a gap closing accompanied by a fermion parity change. This follows directly from combining Eqs.~(\ref{chi-ez}) and (\ref{free-0}) into \cite{SM},
\begin{equation}\label{chi_yy-generic}
\chi_{yy}\propto\sum_{n}\left[\delta(E_{n})\left(\frac{\partial E_{n}}{\partial E_Z}\right)^2+\frac{{\rm sgn}(E_{n})}{2}\frac{\partial^2 E_{n}}{\partial E_Z^2}\right],
\end{equation}
where the presence of the Dirac-delta function anticipates the existence of SSPs when zero energy states emerge.
This expression also explains the correlation between the topological gap and the spin susceptibility. The existence of a sizable topological gap requires the lower (higher) particle-like (hole-like) energy states to exhibit a concave down (up) $E_Z$-dependence with $\partial^2 E_n/\partial E_Z^2<0$ ($\partial^2 E_n/\partial E_Z^2>0$) away from the peaks, where the spin susceptibility is dominated by the contribution involving second derivatives of the  eigenenergies $E_n$. Therefore, positive (negative) energy states with a concave down (up) behavior, favorable to the formation of a sizable topological gap, contribute negatively to the spin susceptibility. As a result, the spin susceptibility tends to decrease when the topological gap increases and vice-versa.

\begin{figure}[t]
\centering
\includegraphics*[width=8.5cm]{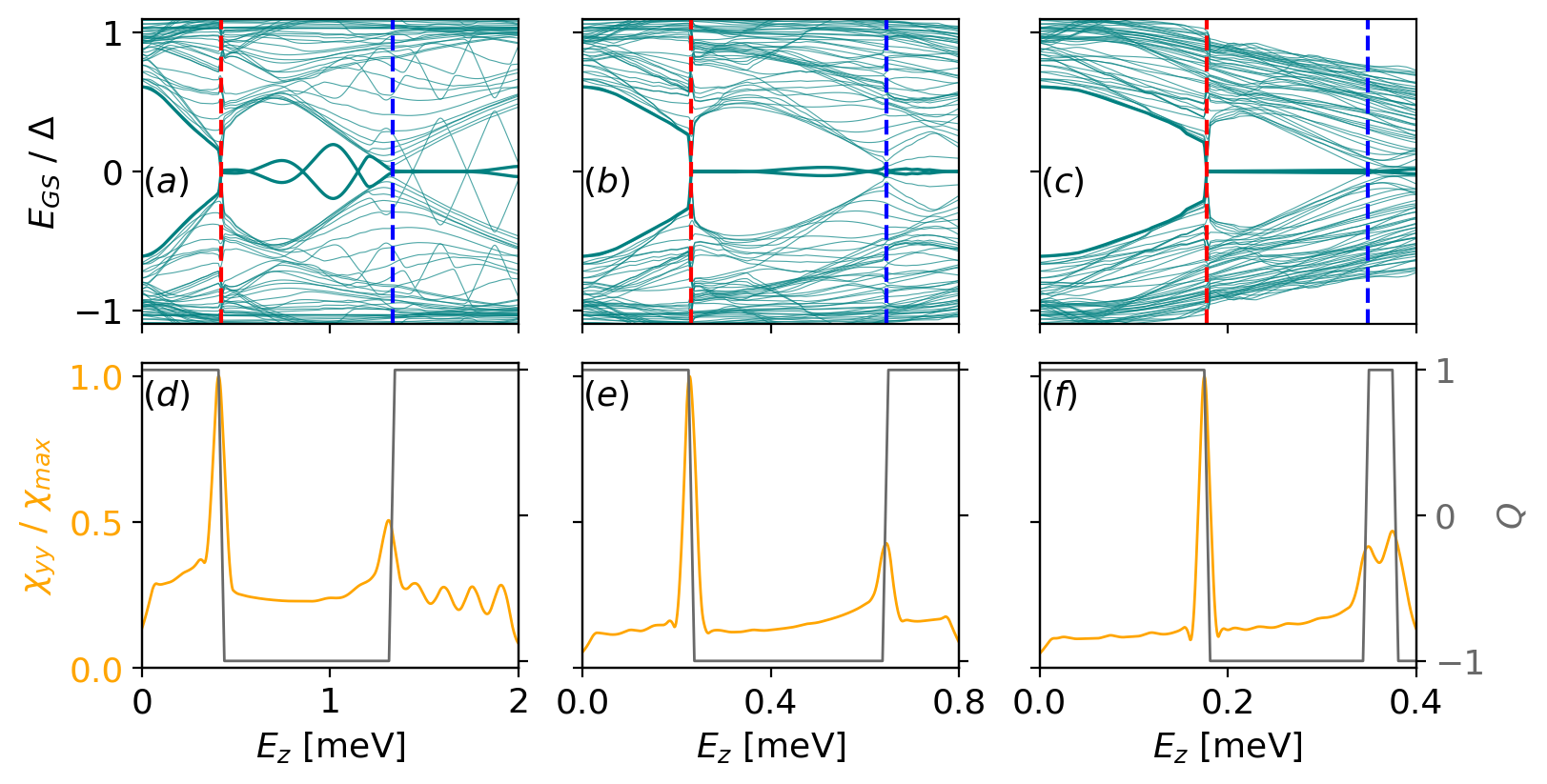}
\caption{Top row: Energy spectrum of a phase-unbiased JJ as a function of the Zeeman energy, $E_Z$, for
$\gamma=1$~nm (a), $\gamma=10^2$~nm (b), and $\gamma=10^4$~nm (c).
Bottom row: Same as in top row but for the topological charge (gray line) and the ratio of the spin susceptibility (orange line) to its maximum value $\chi_{max}$.
The parameters used are: $W_S=450$~nm, $W_N=100$~nm, $a=20$~nm, $\mu_S=1$~meV, and $\mu_N=0.7$~meV. A JJ of length $L=1\mu{\rm m}$ was assumed for the energy spectrum calculations.
}\label{fig4:unbiased}
\end{figure}

\emph{Phase-Unbiased JJs}---In the absence of a phase biasing magnetic flux, the phase difference between the S regions of the JJ self-adjusts in order to minimize the free energy of the system. This enables the system to self-drive into the TS state as $E_z$ is varied, without the need for an external magnetic flux. This is illustrated in Figs.~\ref{fig4:unbiased}(a)-(c), where the ground state spectrum (i.e., the energy spectrum leading to the minimum free energy) is shown as a function of $E_Z$, for different values of the Zeeman field penetration length ($\gamma=1$, $10^2$, and $10^4$~nm, respectively). Right after the first gap closing, indicated by the red dashed line, the JJ transits into the TS state, where MBS emerge inside the topological gap. The second gap closing (marked by the blue dashed line) indicates a transition from the TS to the trivial state. When the Zeeman field is present only in the N region the wavefunctions of the MBSs localized at opposite ends of the junction overlap, producing oscillations of the MBS energy [see \ref{fig4:unbiased}(a)]. However, as the penetration of the Zeeman field into the S regions increases, the MBSs become more localized \cite{SM}, the wavefunction overlap becomes negligible, and the MBS zero energy stabilizes. The topological character of the indicated gap closings is demonstrated by the topological charge $Q$, which as shown in Figs.~\ref{fig4:unbiased}(d)-(f), flips its sign precisely when these closings occur (the TS state is realized when $Q=-1$). The sign changes of the topological charge are accompanied by sharp peaks in the spin susceptibility. Therefore the existence of sharp SSPs can serve as an indication of topological phase transitions also in phase-unbiased planar JJs. When the Zeeman field is present in the N region only, the localization length of the bound states, as well as their penetration in the S regions is relatively large \cite{Fornieri2019:N}, leading to rapid oscillations of the energy states [see Fig.~\ref{fig4:unbiased}(a)]. The oscillatory behavior in the non-topological phase is well captured by the spin susceptibility, as shown in  Fig.~\ref{fig4:unbiased}(d)]. Furthermore, when the JJ is in the TS state, the amplitude of the spin susceptibility serves as a qualitative measure of the size of the topological gap. This can be seen in Figs.~\ref{fig4:unbiased}(d)-(e), where the minimum value of the spin susceptibility between the first two SSPs roughly occurs at a Zeeman field at which the largest topological gap is observed in Figs.~\ref{fig4:unbiased}(a)-(c). Thus, by measuring the spin susceptibility one could not only identify the topological phase transition but also roughly estimate the value of the magnetic field at which TS is realized with the best topological protection.

\begin{figure}[t]
\centering
\includegraphics*[width=8.5cm]{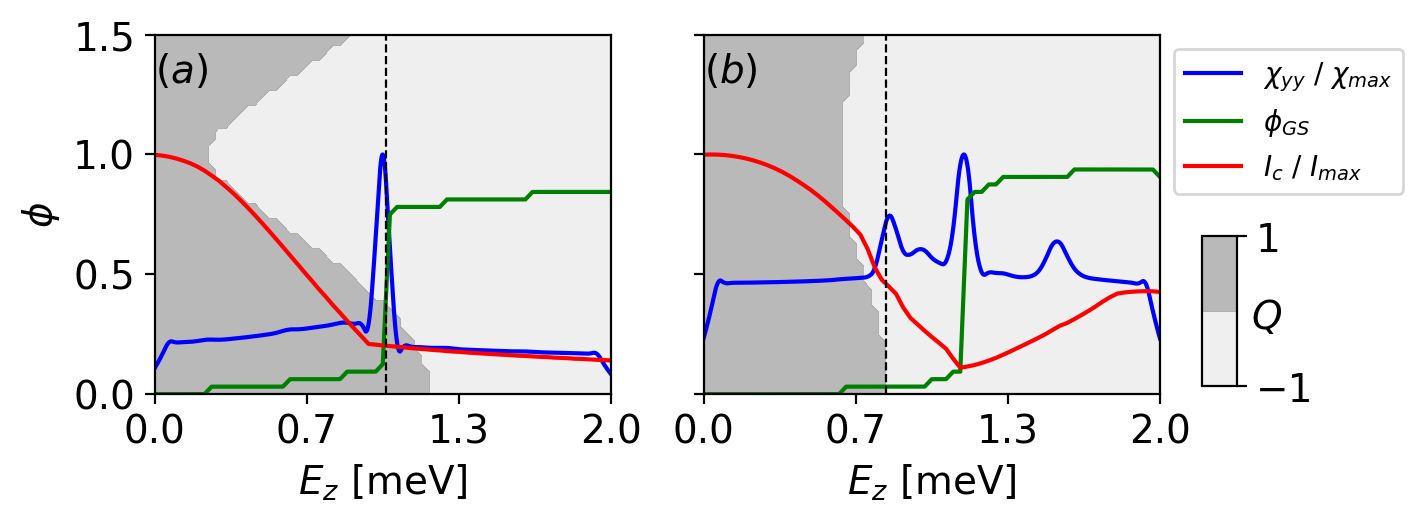}
\caption{Topological phase diagram indicating the Zeeman field and phase dependence of the topological charge. The red, green, and blue lines represent the critical current, ground-state phase, and spin susceptibility, respectively. (a) System parameters: $W_S=300$~nm, $W_N=40$~nm, $a=10$~nm, $\mu_S=1$~meV, and $\mu_N=1$~meV. (b) System parameters: $W_S=100$~nm, $W_N=100$~nm, $a=5$~nm, and $\mu_S=\mu_N=8$~meV.
}\label{fig5:unbiased}
\end{figure}

The existence of minima in the critical current (assumed to be correlated to jumps in the ground-state phase) was proposed as a signature of the topological phase transition \cite{Pientka2017:PRX} and has recently been measured in Al/InAs JJs \cite{Dartiailh2021:PRL}. However, the ground-state phase is obtained by minimizing the free energy, while the critical current results from maximizing the supercurrent. Therefore, although in some cases the ground-state phase jumps appear to occur at the same Zeeman energy at which the critical current minima occur, this is, in general, not necessarily true. Further, the existence of ground-state jumps may not even be accompanied by critical current minima (nor vice versa), especially for junctions with narrow S regions \cite{Setiawan2019:PRB2}. In such cases, the existence of current minima does not necessarily represent a signature of the topological phase transition. This is illustrated in Fig.~\ref{fig5:unbiased}, where the topological phase diagram of a planar JJ, together with the calculated ground-state phase (green line), critical current (red line), and spin susceptibility (blue line) is depicted. The light (dark) gray region with topological charge, $Q=-1$ ($Q=1$), represents the topological (trivial) state. The vertical, dashed line indicates the Zeeman field at which the topological transition occurs. In the example of Fig.~\ref{fig5:unbiased}(a), the ground-sate phase jump and SSP correctly signal the topological phase transition, while the critical current does not exhibit any local minimum. In the example of Fig.~\ref{fig5:unbiased}(b), neither the ground-state phase jump nor the critical current minimum directly represents a transition from the trivial to the topological state. However, a SSP still develops at the topological transition. This demonstrates that compared to previously used signatures such as phase shifts and critical current minima, the existence of SSPs constitutes a more robust indication of topological phase transitions in planar JJs.

The ground-state phase jump in Fig.~\ref{fig5:unbiased}(b) is associated with a transition between different symmetry sub-classes and is accompanied by a SSP with large amplitude. For phase-biased JJs, the SSP with larger amplitudes correspond to topological phase transitions, while smaller SSPs may signal non-topological gap closings. However, for phase-unbiased JJs, the largest SSP results when the ground-state phase jumps \cite{SM}, and other smaller SSPs may occur for Zeeman fields where the topological gap has minima [see Fig.~\ref{fig5:unbiased}(b)]. In any case, as the Zeeman field is increased, the first SSP is always an indication of a transition into the topological regime. Complementing spin susceptibility with phase shift (and/or critical current) measurements can provide a more complete understanding of topological transitions, topological gap size, and transitions between different topological classes. 

\emph{Conclusions}---We have studied the behavior of the spin susceptibility in planar Josephson junctions with Rashba SOC and a magnetic field applied along the junction. Our results show that at topological phase transitions the spin average exhibits jumps leading to sharp peaks in the magnetic field dependence of the spin susceptibility. In the topological regime, the amplitude of the spin susceptibility can serve as a qualitative measure of the relative strength of the topological gap. The signatures on the spin susceptibility appear to be more robust than critical current minima (and ground-state phase jumps) previously used to identify topological transitions. This is particularly relevant for Josephson junctions with narrow leads, where critical current minima and phase shifts may no longer indicate topological transitions while the spin susceptibility peaks remain a good signature. Our findings provide a promising direction in the detection and characterization of topological superconductivity and the topological gap of planar Josephson junctions.

\acknowledgments
\emph{Acknowledgments.} The authors acknowledge support from DARPA Grant No.
DP18AP900007.

\bibliographystyle{apsrev4-2}

%

\end{document}